\documentclass[aps,twocolumn,superscriptaddress,tightenlines]{revtex4}
\usepackage{epsfig,graphicx,times}
\usepackage{amstext}
\usepackage{amsmath}
\usepackage{amssymb}
\usepackage{graphicx}
\usepackage{latexsym}
\usepackage{bm}
\usepackage[colorlinks,citecolor=blue, linkcolor=blue,hyperindex,CJKbookmarks,dvipdfm]{hyperref}

\begin{document}

\title{Nonreciprocal single-photon frequency converter via multiple semi-infinite coupled-resonator waveguides}
\author{Xun-Wei Xu}
\email{davidxu0816@163.com}
\affiliation{Department of Applied Physics, East China Jiaotong University, Nanchang,
330013, China}
\author{Ai-Xi Chen}
\email{aixichen@ecjtu.edu.cn}
\affiliation{Department of Physics, Zhejiang Sci-Tech University, Hangzhou, 310018, China}
\affiliation{Department of Applied Physics, East China Jiaotong University, Nanchang, 330013, China}
\author{Yong Li}
\affiliation{Beijing Computational Science Research Center, Beijing 100193, China}
\affiliation{Synergetic Innovation Center of Quantum Information and Quantum Physics,
University of Science and Technology of China, Hefei 230026, China}
\affiliation{Synergetic Innovation Center for Quantum Effects and Applications, Hunan
Normal University, Changsha 410081, China}
\author{Yu-xi Liu}
\affiliation{Institute of Microelectronics, Tsinghua University, Beijing 100084, China}
\affiliation{Tsinghua National Laboratory for Information Science and Technology
(TNList), Beijing 100084, China}
\date{\today }

\begin{abstract}
We propose to construct a nonreciprocal single-photon frequency converter via multiple
semi-infinite coupled-resonator waveguides (CRWs). We first demonstrate that the
frequency of a single photon can be converted nonreciprocally through two CRWs,
which are coupled indirectly by optomechanical interactions with two
nondegenerate mechanical modes. Based on such nonreciprocity, two different
single-photon circulators are proposed in the T-shaped waveguides consisting
of three semi-infinite CRWs, which are coupled in pairwise
by optomechanical interactions. One circulator is proposed by using two nondegenerate mechanical modes
and the other one is proposed by using three nondegenerate mechanical modes. Nonreciprocal
single-photon frequency conversion is induced by breaking the
time-reversal symmetry, and the optimal conditions for nonreciprocal frequency conversion are
obtained. These proposals can be used to realize nonreciprocal frequency
conversion of single photons in any two distinctive waveguides with different
frequencies and they can allow for dynamic control of the direction of frequency conversion by tuning the phases of external driving lasers, which may have versatile applications in hybrid quantum networks.
\end{abstract}

\maketitle


\section{Introduction}

To build a hybrid quantum network, by harnessing advantages of different systems~%
\cite{WallquistPS09,XiangZLRMP13}, we have to tackle an important problem:
how to integrate different components that don't operate at the same
frequency. One solution is to build a photon frequency converter which converts
the input photons of one frequency into the output photons of another
frequency. Traditionally, photon frequency conversion is demonstrated by
three-wave mixing in second-order nonlinear materials~\cite%
{KumarOL90,JHuangPRL92,RakherNPT10,IkutaNC11,ZaskePRL12,AtesPRL12,AbdoPRL13}
or four-wave mixing in third-order nonlinear materials~\cite%
{McGuinnessPRL10,RadnaevNPy10,FarnesiPRL14}. With the development of circuit
quantum electrodynamics, frequency conversion was even proposed in a single
three-level superconducting quantum circuit by three-wave mixing~\cite%
{YXLiuSR14,YJZhaoArx15} or a single qubit in the ultrastrong coupling regime~%
\cite{KockumArx17}. Moreover, single-photon frequency converters have been
proposed in the one-dimensional (1D) linear waveguide~\cite%
{BradfordPRL12,BradfordPRA12,WBYanSR13} or 1D coupled-resonator waveguides
(CRWs)~\cite{LZhouPRL13,ZHWangPRA14} with a three-level system coupled to
different channels.
Since the mechanical resonators can be coupled to
various electromagnetic fields with distinctively different wavelengths
through radiation pressure (for reviews, see Refs.~\cite%
{KippenbergSci08,MarquardtPhy09,AspelmeyerPT12,AspelmeyerARX13,MetcalfeAPR14}%
), frequency conversion has been demonstrated via two optical cavities with
different frequencies, coupled by a single mechanical resonator via
optomechanical interactions~\cite%
{SafaviNaeiniNJP11,YDWangPRL12,LTianPRL12,JTHillNC12,YLiuPRL13,CDongAP15,LecocqPRL16}. Recently,
the conversion between microwave and optical frequencies has been
implemented in the electro-optomechanical systems~\cite%
{BochmannNPy13,BagciNat14,AndrewsNPy14,RuedaOptica16}.

Besides frequency converters, isolators and circulators are also dispensable
elements in constructing hybrid quantum networks for protecting some
elements from unwanted noises or retracing fields~\cite{JalasNPT13}. It is
well known that, the systems with broken time-reversal symmetry can be used
to construct isolators or circulators. In recent years, as a
non-magnetic strategy, optical nonreciprocity in the coupled cavity modes
with relative phase has drawn more and more attentions, and many different
structures have been proposed theoretically~\cite%
{KochPRA10,HabrakenNJP12,RanzaniNJP14a,XuXWPRA15,RanzaniNJP15a,YPWangSR15a,SliwaPRX15,SchmidtOpt15,MetelmannPRX15,FangKArx15,XXuarX17a,FXSunarX17}
and demonstrated experimentally~\cite{RuesinkNC16a,KFangNPy17a}.

In a recent work, we have proposed a nonreciprocal frequency converter in an
electro-optomechanical system with a microwave mode and an optical mode,
coupled indirectly via two nondegenerate mechanical modes~\cite{XWXuPRA16a}.
Due to the broken time-reversal symmetry, the nonreciprocity is obtained
when the transmission of photons from one mode to the other one is enhanced for
constructive quantum interference while the transmission in the reversal
direction is suppressed with destructive quantum interference. Based on a similar mechanism,
nonreciprocal frequency conversion was explored theoretically~\cite%
{MetelmannarX16a,LTianarx16a,MiriarX16a} and
realized experimentally~\cite{BernierarX16a,PetersonarX17,BarzanjeharX17} in many different systems.

In this paper, we propose a nonreciprocal single-photon frequency converter,
consisting of two or three 1D semi-infinite CRWs with different frequencies, which are coupled
indirectly by nondegenerate mechanical modes via optomechanical
interactions. In quantum networks, this system can also be viewed as quantum
channels (1D semi-infinite CRWs) connected by a quantum node (the
optomechanical systems~\cite{XWXuPRA16a}). Different from the previous
studies on nonreciprocal frequency conversion~\cite%
{XWXuPRA16a,MetelmannarX16a,LTianarx16a,MiriarX16a,BernierarX16a}, we consider the
dispersion relations of the quantum channels, which play an important role in
single-photon frequency conversion. Also unlike the previous studies on
single-photon nonreciprocity in 1D CRWs~\cite{XWXuPRA17,SYangArx09,XQLiPRA15}, in this work, the frequencies of
the CRWs are very different and they can not be coupled together directly.
The addition of optomechanical systems (or mechanical modes) to the frequency converter offers the possibility to enable nonreciprocal frequency transduction between two CRWs with distinctively different frequencies and allows for
dynamic control of the direction of frequency conversion by tuning the phases of external driving lasers.

The paper is organized as follows: In Sec.~II, we propose a single-photon
frequency converter using two CRWs, coupled indirectly by two nondegenerate
mechanical modes via optomechanical interactions. In Secs.~III and IV, two
different single-photon circulators are proposed in the T-shaped waveguides consisting of three semi-infinite CRWs, which are mutually coupled by optomechanical interactions. One circulator uses two
nondegenerate mechanical modes and the other one uses three nondegenerate mechanical
modes. Finally, we summarize our results in Sec.~V.

\section{Nonreciprocal single-photon frequency converter}

\subsection{Theoretical model and scattering matrix}

\begin{figure}[tbp]
\includegraphics[bb=5 478 591 701, width=8.5 cm, clip]{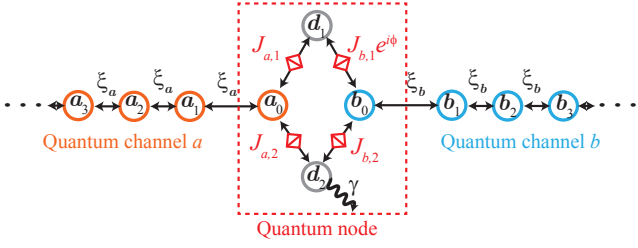}
\caption{(Color online) Schematic diagram of a waveguide consisting of two
semi-infinite CRWs ($a_{j}$ and $b_{j}$ for $j\geq 0$) with different
frequencies (e.g., one is optical CRW and the other one is microwave CRW)
coupled indirectly by the two mechanical modes ($d_{1}$ and $d_{2}$).}
\label{fig1}
\end{figure}

As schematically shown in Fig.~\ref{fig1}, a waveguide consists of two semi-infinite
coupled-resonator waveguides (CRWs) with different frequencies (e.g., one is
optical CRW and the other one is microwave CRW), in which both end side cavities are coupled to
two mechanical modes via optomechanical interactions. The
semi-infinite CRWs, as quantum channels for single-photon transmission, are made by infinite identical single-mode cavities, which are coupled to each other
through coherent hopping of photons between neighboring cavities~\cite%
{WallraffNat04,HartmannNP06,NotomiNPT08,CastellanosAPL07}. The two
end side cavities ($a_{0}$ and $b_{0}$), coupled indirectly by the two
mechanical modes ($d_{1}$ and $d_{2}$), are served as a quantum node for
single-photon frequency conversion. The total system can be described by the
Hamiltonian
\begin{equation}\label{eq1}
H_{0}=\sum_{l=a,b}H_{l}+H_{m}+H_{\mathrm{int}}
\end{equation}%
with the Hamiltonian $\sum_{l=a,b}H_{l}$ for the two CRWs
\begin{equation}\label{eq2}
H_{l}=\sum_{j=0}^{+\infty }\left[ \omega _{l}l_{j}^{\dag }l_{j}-\xi
_{l}\left( l_{j}^{\dag }l_{j+1}+\mathrm{H.c.}\right) \right] ,
\end{equation}%
the Hamiltonian $H_{m}$ for the mechanical modes%
\begin{equation}
H_{m}=\omega _{1}d_{1}^{\dag }d_{1}+\omega _{2}d_{2}^{\dag }d_{2},
\end{equation}%
and the interaction terms $H_{\mathrm{int}}$ for single-photon frequency
conversion%
\begin{eqnarray}
H_{\mathrm{int}} &=&\left( g_{a,1}a_{0}^{\dag }a_{0}+g_{b,1}b_{0}^{\dag
}b_{0}\right) \left( d_{1}+d_{1}^{\dag }\right)  \notag \\
&&+\left( g_{a,2}a_{0}^{\dag }a_{0}+g_{b,2}b_{0}^{\dag }b_{0}\right) \left(
d_{2}+d_{2}^{\dag }\right)  \notag \\
&&+\sum_{l=a,b}\sum_{i=1}^{2}\left( l_{0}\Omega _{l,i}e^{i\omega _{l,i}t}+%
\mathrm{H.c.}\right) ,
\end{eqnarray}%
where $l_{j}$ ($l_{j}^{\dag }$, $l=a,b$) is the bosonic annihilation
(creation) operator of the $j$th cavity with the same resonant frequency $%
\omega _{l}$ and the same coupling strength $\xi _{l}$ between two nearest neighboring
cavities in the CRW-$l$. $\omega _{i}$ ($i=1,2$) is the resonant
frequency of the mechanical mode with the bosonic annihilation (creation)
operator $d_{i}$ ($d_{i}^{\dag }$). $g_{l,i}$ is the optomechanical coupling
strength between cavity $l_{0}$ ($l_{0}=a_{0},b_{0}$) and mechanical mode $d_{i}$ ($%
d_{i}=d_{1},d_{2}$). The cavity $a_{0}$ ($b_{0}$) is driven by a two-tone laser
at frequencies $\omega _{a,1}=\omega _{a}-\omega _{1}+\Delta _{a,1}$ and $%
\omega _{a,2}=\omega _{a}-\omega _{2}+\Delta _{a,2}$ ($\omega _{b,1}=\omega
_{b}-\omega _{1}+\Delta _{b,1}$ and $\omega _{b,2}=\omega _{b}-\omega
_{2}+\Delta _{b,2}$) with amplitudes $\Omega _{a,1}$ and $\Omega _{a,2}$ ($%
\Omega _{b,1}$ and $\Omega _{b,2}$). For simplicity, we assume that $\Delta _{1}\equiv \Delta _{a,1}=\Delta _{b,1}$ and  $\Delta _{2}\equiv \Delta _{a,2}=\Delta _{b,2}$. Thus the operators for the cavity modes
can be rewritten as the sum of the quantum fluctuation operators and
classical mean values, i.e., $a_{j}\rightarrow a_{j}+\sum_{i=1}^{2}\alpha^{a}
_{j,i}e^{-i\omega_{a,i}t}$ and $b_{j}\rightarrow
b_{j}+\sum_{i=1}^{2}\alpha^{b} _{j,i}e^{-i\omega _{b,i}t}$, where $a_{j}$ and $b_{j}$ on the right side of the arrow symbols
describe the quantum fluctuation operators of the cavity modes, and the
classical amplitude $\alpha^{l} _{j,i}$ is determined by the amplitudes $%
\Omega _{l,i}$, the frequency $\omega _{l,i}$, the damping rates $\kappa
_{a,j}$ and $\kappa _{b,j}$ of the cavities and the damping rates $\gamma
_{1}$ and $\gamma _{2}$ of the mechanical modes.

To obtain a linearized Hamiltonian, we assume that the external driving is
strong, i.e. $|\alpha^{l} _{0,i}|\gg 1$, the system works in the
resolved-sideband limit with respect to both mechanical modes, i.e. $\min
\left\{ \omega _{1},\omega _{2}\right\} \gg \max \left\{ \kappa
_{a,j},\kappa _{b,j}\right\} $, and the two mechanical modes are well
separated in frequency, i.e. $\min \left\{ \omega _{1},\omega
_{2},\left\vert \omega _{1}-\omega _{2}\right\vert \right\} \gg \max \left\{
|g_{l,i}\alpha^{l} _{0,i}|,\gamma _{1},\gamma _{2}\right\} $. After making
the standard linearization under the rotating-wave approximation, in the
rotating reference frame with respect to $H_{\mathrm{rot}}=\sum_{l=a,b}\sum_{j=0}^{+%
\infty }\omega _{l}l_{j}^{\dag }l_{j}+\sum_{i=1,2}\left( \omega _{i}-\Delta
_{i}\right) d_{i}^{\dag }d_{i}$, the linearized Hamiltonian with time-independent terms becomes
\begin{equation}\label{eq5}
H_{\mathrm{fc}}=\sum_{l=a,b}H_{l}+H_{m}+H_{\mathrm{int}},
\end{equation}%
where $H_{l}$, $H_{m}$, and $H_{\mathrm{int}}$ are replaced by
\begin{equation}\label{eq6}
H_{l}=-\xi _{l}\sum_{j=0}^{+\infty }\left( l_{j}^{\dag }l_{j+1}+\mathrm{H.c.}%
\right) ,
\end{equation}%
\begin{equation}\label{eq7}
H_{m}=\Delta _{1}d_{1}^{\dag }d_{1}+\Delta _{2}d_{2}^{\dag }d_{2},
\end{equation}%
\begin{eqnarray}\label{eq8}
H_{\mathrm{int}} &=&J_{a,1}(a_{0}^{\dag }d_{1}+a_{0}d_{1}^{\dag })  \notag \\
&&+J_{b,1}(e^{-i\phi }b_{0}^{\dag }d_{1}+e^{i\phi }b_{0}d_{1}^{\dag })
\notag \\
&&+J_{a,2}(a_{0}^{\dag }d_{2}+a_{0}d_{2}^{\dag })  \notag \\
&&+J_{b,2}(b_{0}^{\dag }d_{2}+b_{0}d_{2}^{\dag }).
\end{eqnarray}%
Here $J_{l,i}e^{i\phi _{l,i}}=g_{l,i}\alpha^{l} _{0,i}$ is the effective
optomechanical coupling strength between the cavity $l_{0}$ ($%
l_{0}=a_{0},b_{0}$) and mechanical mode $d_{i}$ ($d_{i}=d_{1},d_{2}$) with
real strength $J_{l,i}=\left\vert g_{l,i}\alpha^{l} _{0,i}\right\vert $ and
phase $\phi _{l,i}$. As only the total phase $\phi =\phi
_{a,1}+\phi _{a,2}+\phi _{b,1}+\phi _{b,2}$ has physical effects,
without loss of generality, $\phi $ is only kept in the terms of $%
b_{0}d_{1}^{\dag }$ and $b_{0}^{\dag }d_{1}$ in Eq.~(\ref{eq8}) and the following
derivation. It should be noted that $\phi $ and $J_{l,i}$ are dynamically tunable parameters, which can be controlled by tuning the strengths and phases of the external driving fields. The time-reversal symmetry of the whole
system is broken when we choose the phase $\phi \neq n\pi $ ($n$ is an
integer). As we will show later, the direction of frequency conversion can be controlled dynamically by tuning the value of the total phase $\phi$.

In this paper, we assume that the damping rates of the cavities in the CRWs
are much smaller than the coupling strength between two nearest neighboring
cavities and the effective optomechanical coupling strength, i.e. $%
\left\{\xi _{l},J_{l,i}\right\}\gg \max \left\{ \kappa _{a,j},\kappa
_{b,j}\right\} $, so that we can only consider the coherent scattering in
the CRWs. Moreover, we assume that $\left\{\xi
_{l},J_{l,i},\gamma _{2}\right\}\gg \gamma _{1}$, so that $\gamma _{1}$ can be neglected
in the following calculations and the Hamiltonian for two mechanical modes
with $\gamma \equiv \gamma _{2}$ is described by
\begin{equation}
H_{m}=\Delta _{1}d_{1}^{\dag }d_{1}+\left( \Delta _{2}-i\gamma \right)
d_{2}^{\dag }d_{2}.
\end{equation}%
The mechanical damping rate $\gamma $ can be controlled by coupling the
mechanical mode to an auxiliary cavity~\cite{WilsonRaePRL07,MarquardtPRL07,LiYPRB08,XWXuPRA15}, and a suitable mechanical damping
is another crucial condition to obtain desired nonreciprocal single-photon
frequency conversion in this model~\cite{XWXuPRA17}.

To derive the sacttering matrix between different CRWs, we consider the
stationary eigenstate of a single photon in the whole system as
\begin{eqnarray}\label{eq10}
\left\vert E\right\rangle  &=&\sum_{j=0}^{+\infty }\left[ u_{a}\left(
j\right) a_{j}^{\dag }\left\vert 0\right\rangle +u_{b}\left( j\right)
b_{j}^{\dag }\left\vert 0\right\rangle \right]   \notag  \\
&&+u_{d1}d_{1}^{\dag }\left\vert 0\right\rangle +u_{d2}d_{2}^{\dag
}\left\vert 0\right\rangle ,
\end{eqnarray}%
where $\left\vert 0\right\rangle $ indicates the vacuum state of the whole
system, $u_{l}\left( j\right) $ denotes the probability amplitude in the
state with a single photon in the $j$th cavity of the CRW-$l$, and $u_{d1}$ ($%
u_{d2}$) denotes the probability amplitude with a single phonon in the
mechanical mode $d_{1}$ ($d_{2}$). The dispersion relation of the
semi-infinite CRW-$l$ in the rotating reference frame is given by~\cite{LZhouPRL13}
\begin{equation}\label{eq11}
E_{l}=-2\xi _{l}\cos k_{l},\quad 0<k_{l}<\pi ,
\end{equation}%
where $E_{l}$ is the energy and $k_{l}$ is the wave number of the
single photon in the CRW-$l$. Without loss of generality, we assume that $\xi _{l}>0$. Substituting the stationary eigenstate in
Eq.~(\ref{eq10}) and the Hamiltonian in Eq.~(\ref{eq5}) into the eigenequation $H_{\mathrm{fc}%
}\left\vert E\right\rangle =E\left\vert E\right\rangle $, we can obtain the
coupled equations for the probability amplitudes as%
\begin{equation}\label{eq12}
J_{a,1}u_{d1}+J_{a,2}u_{d2}-\xi _{a}u_{a}\left( 1\right) =Eu_{a}\left(
0\right) ,
\end{equation}%
\begin{equation}
J_{b,1}e^{-i\phi }u_{d1}+J_{b,2}u_{d2}-\xi _{b}u_{b}\left( 1\right)
=Eu_{b}\left( 0\right) ,
\end{equation}%
\begin{equation}
J_{a,1}u_{a}\left( 0\right) +J_{b,1}e^{i\phi }u_{b}\left( 0\right) =\left(
E-\Delta _{1}\right) u_{d1},
\end{equation}%
\begin{equation}\label{eq15}
J_{a,2}u_{a}\left( 0\right) +J_{b,2}u_{b}\left( 0\right) =\left( E-\Delta
_{2}+i\gamma \right) u_{d2},
\end{equation}%
\begin{equation}\label{eq16}
Eu_{l}\left( j\right) +\xi _{l}u_{l}\left( j+1\right) +\xi _{l}u_{l}\left(
j-1\right) =0
\end{equation}%
with $j>0$ and $l=a,b$.

If a single photon with energy $E$ is incident from the infinity side of CRW-%
$l$, the photon-phonon interactions in the quantum node will result in
photon sacttering between different CRWs or photon absorbtion by the dissipative
mechanical mode. The general expressions of the probability amplitudes in
the CRWs ($j\geq 0$) are given by
\begin{equation}\label{eq17}
u_{l}\left( j\right) =e^{-ik_{l}j}+s_{ll}e^{ik_{l}j},
\end{equation}%
\begin{equation}\label{eq18}
u_{l^{\prime }}\left( j\right) =s_{l^{\prime }l}e^{ik_{l^{\prime }}j},
\end{equation}%
where $s_{l^{\prime }l}$ denotes the single-photon scattering amplitude from
CRW-$l$ to CRW-$l^{\prime }$ ($l,l^{\prime }=a,b$). Substituting
Eqs.~(\ref{eq17}) and (\ref{eq18}) into Eqs.~(\ref{eq12})-(\ref{eq16}), then we obtain the scattering matrix as%
\begin{equation}
S=\left(
\begin{array}{cc}
s_{aa} & s_{ab} \\
s_{ba} & s_{bb}%
\end{array}%
\right) ,
\end{equation}%
where
\begin{equation}\label{eq20}
s_{aa}=D^{-1}\left[ J_{ab}J_{ba}-\left( \xi _{a}e^{ik_{a}}+\Delta %
_{a}\right) \left( \xi _{b}e^{-ik_{b}}+\Delta _{b}\right) \right]
,
\end{equation}%
\begin{equation}\label{eq21}
s_{ba}=i2D^{-1}J_{ba}\xi _{a}\sin k_{a},
\end{equation}%
\begin{equation}\label{eq22}
s_{ab}=i2D^{-1}J_{ab}\xi _{b}\sin k_{b},
\end{equation}%
\begin{equation}
s_{bb}=D^{-1}\left[ J_{ab}J_{ba}-\left( \xi _{a}e^{-ik_{a}}+\Delta
_{a}\right) \left( \xi _{b}e^{ik_{b}}+\Delta _{b}\right) \right],
\end{equation}%
\begin{equation}\label{eq24}
D=\left( \xi _{a}e^{-ik_{a}}+\Delta _{a}\right) \left( \xi
_{b}e^{-ik_{b}}+\Delta _{b}\right) -J_{ab}J_{ba}
\end{equation}%
with the effective coupling strengths $J_{ll^{\prime }}$ and frequency
shifts $\Delta _{l}$ induced by the two mechanical modes defined by
\begin{equation}\label{eq25}
J_{ab}\equiv\frac{J_{a,1}J_{b,1}e^{i\phi }}{\left( E-\Delta _{1}\right) }+\frac{%
J_{a,2}J_{b,2}}{\left( E-\Delta _{2}+i\gamma \right) },
\end{equation}%
\begin{equation}\label{eq26}
J_{ba}\equiv\frac{J_{a,1}J_{b,1}e^{-i\phi }}{\left( E-\Delta _{1}\right) }+\frac{%
J_{a,2}J_{b,2}}{\left( E-\Delta _{2}+i\gamma \right) },
\end{equation}%
\begin{equation}\label{eq27}
\Delta _{a}\equiv\frac{\left( J_{a,1}\right) ^{2}}{\left( E-\Delta _{1}\right) }+%
\frac{\left( J_{a,2}\right) ^{2}}{\left( E-\Delta _{2}+i\gamma \right) },
\end{equation}%
\begin{equation}\label{eq28}
\Delta _{b}\equiv\frac{\left( J_{b,1}\right) ^{2}}{\left( E-\Delta _{1}\right) }+%
\frac{\left( J_{b,2}\right) ^{2}}{\left( E-\Delta _{2}+i\gamma \right) }.
\end{equation}%
To quantify nonreciprocity conversion, we define the scattering flows of the
single photons from CRW-$l$ to CRW-$l^{\prime }$ as~\cite{ZHWangPRA14}
\begin{equation}
I_{l^{\prime }l}=|s_{l^{\prime }l}|^{2}\frac{\xi _{l^{\prime }}\sin
k_{l^{\prime }}}{\xi _{l}\sin k_{l}},
\end{equation}%
where $\xi _{l}\sin k_{l}$ ($\xi _{l^{\prime }}\sin
k_{l^{\prime }}$) is the group velocity in the CRW-$l$ (CRW-$l^{\prime }$).
In our model, $I_{ba}\neq I_{ab}$ implies the appearance of nonreciprocal
single-photon frequency conversion, and the perfect nonreciprocal
single-photon frequency conversion is obtained when $I_{ba}=1$ and $I_{ab}=0$,
or $I_{ba}=0$ and $I_{ab}=1$.

\subsection{Nonreciprocal single-photon frequency converter}

\begin{figure}[tbp]
\includegraphics[bb=72 220 550 605, width=8.5 cm, clip]{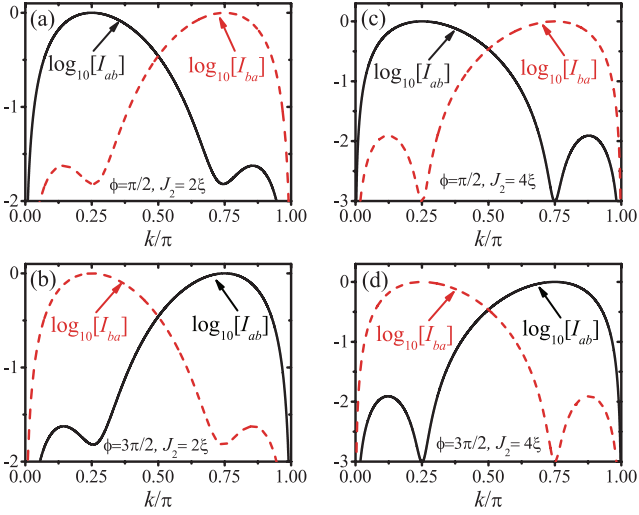}
\caption{(Color online) Scattering flows $\log_{10} [I_{ab}]$ (black solid
curves) and $\log_{10} [I_{ba}]$ (red dashed curves) are shown as functions
of the wave number $k/\protect\pi$ for: (a) $\protect\phi=\protect\pi/2$ and
$J_{2}=2\protect\xi$, (b) $\protect\phi=3\protect\pi/2$ and $J_{2}=2\protect%
\xi$, (c) $\protect\phi=\protect\pi/2$ and $J_{2}=4\protect\xi$, (d) $%
\protect\phi=3\protect\pi/2$ and $J_{2}=4\protect\xi$. The other parameters
are $J_{1}=\protect\xi $, $\Delta _{1}=\Delta _{2}=0$, and $\protect\gamma$
is obtained from Eq.~(\ref{eq32}).}
\label{fig2}
\end{figure}

\begin{figure}[tbp]
\includegraphics[bb=58 139 553 561, width=8.5 cm, clip]{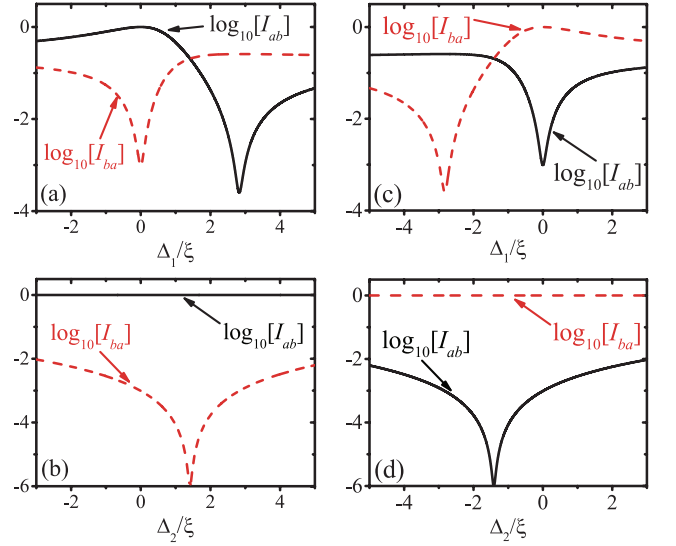}
\caption{(Color online) Scattering flows $\log_{10} [I_{ab}]$ (black solid
curves) and $\log_{10} [I_{ba}]$ (red dashed curves) are plotted as
functions of $\Delta _{1}/\protect\xi$ in (a) and (c), and as functions of $%
\Delta _{2}/\protect\xi$ in (b) and (d) with wave numbers: (a) and (b) $k=%
\protect\pi/4$, (c) and (d) $k=3\protect\pi/4$. The other parameters are $%
\protect\phi=\protect\pi/2$, $J_{1}=\protect\xi $, $J_{2}=4\protect\xi$, and $%
\protect\gamma$ is obtained from Eq.~(\ref{eq32}).}
\label{fig3}
\end{figure}

Before the numerical calculations of the scattering flows $I_{ab}$ and $%
I_{ba}$, it is instructive to find the optimal conditions for nonreciprocal
single-photon frequency conversion analytically. For simplicity, we assume
that the detunings $\Delta _{1}=\Delta _{2}=0 $, the two CRWs have
the same parameters (i.e., $\xi \equiv \xi _{a}=\xi _{b} $, and $k\equiv
k_{a}=k_{b}$) and they are symmetrically coupled to the two mechanical modes
($J_{1}\equiv J_{a,1}=J_{b,1}$, $J_{2}\equiv J_{a,2}=J_{b,2} $) with $%
J_{1}=\xi $. Under the condition that $\gamma \gg \xi $, the optimal
conditions for nonreciprocal single-photon frequency conversion obtained
from Eqs.~(\ref{eq20})-(\ref{eq24}) are
\begin{equation}\label{eq30}
\phi \approx \frac{\pi }{2}\quad \mathrm{or}\quad \frac{3\pi }{2},
\end{equation}%
\begin{equation}\label{eq31}
k\approx \frac{\pi }{4}\quad \mathrm{or}\quad \frac{3\pi }{4},
\end{equation}%
\begin{equation}\label{eq32}
\frac{\gamma }{\xi }\approx \sqrt{2\left( \frac{J_{2}}{\xi }\right) ^{4}+2}.
\end{equation}%
In order to satisfy the condition $\gamma \gg \xi $, we should choose $J_{2}\gg \xi $ in Eq.~(\ref{eq32}).

Scattering flows $I_{ab}$ (black solid curve) and $I_{ba}$ (red dashed
curve) as functions of the wave number $k/\pi $ are shown in Fig.~\ref{fig2}.
The optimal nonreciprocity appears around the point $k\approx \pi/4$ and $%
3\pi/4$ for $\phi \approx \pi/2$ or $3\pi/2$, which exhibits good agreement
with the analytical result shown in Eqs.~(\ref{eq30}) and (\ref{eq31}).
Specifically, when $\phi = \pi/2$, in Figs.~\ref{fig2}(a) and \ref{fig2}(c), we show the reciprocal transmission from CRW-$b$ to CRW-$a$ (CRW-$a$ to CRW-$b$) at $k\approx \pi/4$ ($k\approx 3\pi/4$).
In contrast, when $\phi = 3\pi/2$, we see the reciprocal transmission from CRW-$a$ to CRW-$b$ (CRW-$b$ to CRW-$a$) at $k\approx \pi/4$ ($k\approx 3\pi/4$).
These imply that we can reverse the direction of frequency conversion by tuning the phase from $\phi=\pi/2$ to $\phi=3\pi/2$.
Scattering flows $I_{ab}$ and $I_{ba}$ for different $J_{2}$ are shown in Figs.~\ref{fig2}(a) and \ref{fig2}%
(c) [or Figs.~\ref{fig2}(b) and \ref{fig2}(d)], which demonstrate that the
nonreciprocity of the system improves dramatically if we take a larger value of $J_{2}/\xi$ (as well as $%
\gamma/\xi$).

The optimal conditions for nonreciprocal single-photon frequency conversion
given in Eqs.~(\ref{eq30})-(\ref{eq32}) are only applicable for zero frequency detunings $\Delta _{1}=\Delta
_{2}=0 $. The following discussions based on
numerical calculations will show the effects of the frequency
detunings ($\Delta _{1}$ and $\Delta_{2}$) on frequency conversion. Scattering flows $I_{ab}$ (black solid curve) and $%
I_{ba}$ (red dashed curve) as functions of the detunings $\Delta _{1}/\xi $
and $\Delta _{2}/\xi $ are shown in Fig.~\ref{fig3}. From these figures, we
can see two interesting phenomena. (i) Besides $\Delta _{1}=0$, there is
another optimal detuning $\Delta _{1}= 2\sqrt{2}\xi$ ($\Delta _{1}= -2\sqrt{2%
}\xi$) for observing nonreciprocity in the oppose direction with the wave number $k=\pi/4$ ($k=3\pi/4$).
Thus we can change the direction of the scattering flows from $b\rightarrow a
$ to $a\rightarrow b$ (from $a\rightarrow b $ to $b\rightarrow a$) by tuning the detuning from $\Delta _{1}=0$ to $\Delta
_{1}= 2\sqrt{2}\xi$ ($\Delta _{1}=- 2\sqrt{2}\xi$). This phenomenon can be
simply understood by plugging $\Delta _{1}= 2\sqrt{2}\xi$ and $k=\pi/4$ ($%
\Delta _{1}= -2\sqrt{2}\xi$ and $k=3\pi/4$) into Eqs.~(\ref{eq21}) and (\ref{eq22}), then we
obtain $I_{ab} \approx 0$ and $I_{ba} \approx 0.258$ ($I_{ba} \approx 0$ and
$I_{ab} \approx 0.258$) for $\phi=\pi/2$. (ii) We can improve the
nonreciprocity by taking $\Delta _{2}=\sqrt{2}\xi$ for $k=\pi/4$ ($\Delta
_{2}=-\sqrt{2}\xi$ for $k=3\pi/4$). This phenomenon corresponds to the condition $%
E=\Delta_{2}$ for $k=\pi/4$ (or $k=3\pi/4$) in Eqs.~(\ref{eq25})-(\ref{eq28}), so that the
optimal phase is $\phi=\pi/2$ (or $\phi=3\pi/2$).

\section{Single-photon circulator in T-shaped waveguide}

\subsection{Theoretical model and scattering matrix}

\begin{figure}[tbp]
\includegraphics[bb=5 242 588 701, width=8.5 cm, clip]{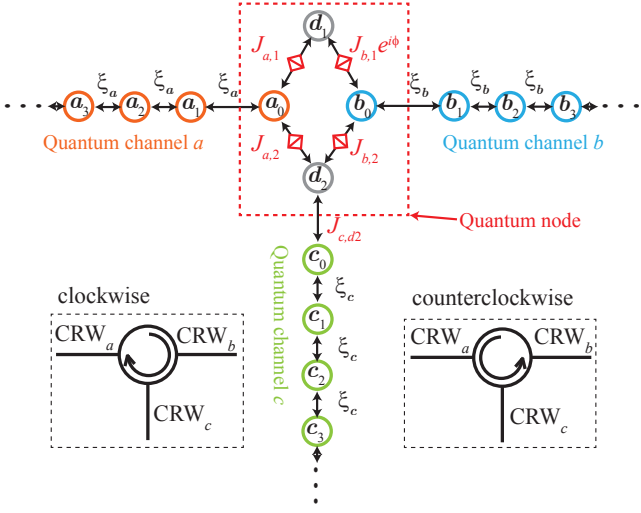}
\caption{(Color online) Schematic diagram of a T-shaped waveguide consisting
of three semi-infinite CRWs ($a_{j}$, $b_{j}$ and $c_{j}$ for $j\geq 0$)
coupled indirectly by two mechanical modes ($d_{1}$ and $d_{2}$).}
\label{fig4}
\end{figure}

Based on the nonreciprocal single-photon frequency conversion discussed in Sec.~II,
we propose a three-port single-photon circulator in a dissipation-free
T-shaped waveguide, i.e. $\gamma =0$, as schematically shown in Fig.~\ref{fig4}, which is made
up by coupling an additional semi-infinite CRW (CRW-$c$) to
the mechanical mode $d_{2}$ in Fig.~\ref{fig1} through optomechanical interaction. The T-shaped
waveguide can be described by the Hamiltonian
\begin{equation}
H_{\mathrm{T,I}}=H_{0}+H_{c}+H_{c,\mathrm{int}},
\end{equation}%
where $H_{0}$ is given in Eq.~(\ref{eq1}), and the two additional terms $H_{c}$ and $%
H_{c,\mathrm{int}}$ are%
\begin{equation}
H_{c}=\sum_{j=0}^{+\infty }\left[ \omega _{c}c_{j}^{\dag }c_{j}-\xi
_{c}\left( c_{j}^{\dag }c_{j+1}+\mathrm{H.c.}\right) \right] ,
\end{equation}%
\begin{equation}
H_{c,\mathrm{int}} =g_{c,2}c_{0}^{\dag }c_{0}\left( d_{2}+d_{2}^{\dag
}\right)+\left( c_{0}\Omega _{c,2}e^{i\omega _{c,2}t}+\mathrm{H.c.}\right).
\end{equation}%
Here $c_{j}$ ($c_{j}^{\dag }$) is the bosonic annihilation (creation)
operator of the $j$th cavity with the same resonant frequency $\omega _{c}$
and the same coupling strength $\xi_{c}$ between two nearest neighboring
cavities in the CRW-$c$. $g_{c,2}$ is the optomechanical coupling strength
between the cavity $c_{0}$ and the mechanical mode $d_{2}$. Cavity $c_{0}$ is driven
by a laser at frequency $\omega _{c,2}=\omega _{c}-\omega _{2}+\Delta _{c,2}$
with $\Delta _{2}=\Delta _{c,2}$ and amplitude $\Omega _{c,2}$. Similarly, the operator for the cavity $c_{j}
$ can also be rewritten as the sum of its quantum fluctuation operator and
classical mean value as $c_{j}\rightarrow c_{j}+\alpha ^{c}_{j,2}e^{-i\omega
_{c,2}t}$, where the classical amplitude $\alpha^{c} _{j,2}$ is determined
by the amplitude $\Omega _{c,2}$, the frequency $\omega _{c,2}$, the
damping rates $\kappa _{a,j}$, $\kappa _{b,j}$ and $\kappa _{c,j}$ of the
cavities and the mechanical damping rates $\gamma _{1}$ and $\gamma _{2}$.

To obtain a linearized Hamiltonian, we assume that the external driving is
strong, i.e. $|\alpha^{l} _{0,i}|\gg 1$, the system works in the
resolved-sideband limit with respect to both mechanical modes, i.e. $\min
\left\{ \omega _{1},\omega _{2}\right\} \gg \max \left\{ \kappa
_{a,j},\kappa _{b,j},\kappa _{c,j}\right\} $, and the two mechanical modes
are well separated in frequency, i.e. $\min \left\{ \omega _{1},\omega
_{2},\left\vert \omega _{1}-\omega _{2}\right\vert \right\} \gg \max \left\{
|g_{l,i}\alpha^{l} _{0,i}|,\gamma _{1},\gamma _{2}\right\} $. After making
the standard linearization under the rotating-wave approximation, in the
rotating reference frame with respect to $H_{\mathrm{rot}}=\sum_{l=a,b,c}\sum_{j=0}^{+%
\infty }\omega _{l}l_{j}^{\dag }l_{j}+\sum_{i=1,2}\left( \omega _{i}-\Delta
_{i}\right) d_{i}^{\dag }d_{i}$, the linearized Hamiltonian for the T-shaped
waveguide is described by%
\begin{equation}
H_{\mathrm{cir,I}}=\sum_{l=a,b,c}H_{l}+H_{m}+H_{\mathrm{int,I}},
\end{equation}%
where $H_{l}$ and $H_{m}$ have been given in Eqs.~(\ref{eq6}) and (\ref{eq7}), and the
interaction term $H_{\mathrm{int,I}}$ is given by
\begin{eqnarray}
H_{\mathrm{int,I}} &=&J_{a,1}\left( a_{0}^{\dag }d_{1}+a_{0}d_{1}^{\dag
}\right)  \notag \\
&&+J_{b,1}\left( e^{-i\phi }b_{0}^{\dag }d_{1}+e^{i\phi }b_{0}d_{1}^{\dag
}\right)  \notag \\
&&+J_{a,2}\left( a_{0}^{\dag }d_{2}+a_{0}d_{2}^{\dag }\right)  \notag \\
&&+J_{b,2}\left( b_{0}^{\dag }d_{2}+b_{0}d_{2}^{\dag }\right)  \notag \\
&&+J_{c,2}\left( c_{0}^{\dag }d_{2}+c_{0}d_{2}^{\dag }\right) .
\end{eqnarray}%
Here $J_{c,2}=\left\vert g_{c,2}\alpha^{c} _{0,2}\right\vert $ is the
effective optomechanical coupling strength between the cavity $c_{0}$ and
mechanical mode $d_{2}$. We assume that the damping rates $%
\kappa _{l,j}$ of the cavities in the CRWs and the damping rates $\gamma _{i}
$ of the mechanical modes are much smaller than the coupling strengths
between two nearest neighboring cavities and the effective optomechanical
coupling strengths, i.e. $\left\{\xi_{l},J_{l,i}\right\}\gg \max \left\{
\kappa _{l,j},\gamma _{i}\right\} $, so that we can only consider the
coherent scattering in the CRWs.

The stationary eigenstate of a single-photon scattering in the T-shaped
waveguide is given by
\begin{equation}
\left\vert E\right\rangle =\sum_{l=a,b,c}\sum_{j=0}^{+\infty }u_{l}\left(
j\right) l_{j}^{\dag }\left\vert 0\right\rangle +u_{d1}d_{1}^{\dag
}\left\vert 0\right\rangle +u_{d2}d_{2}^{\dag }\left\vert 0\right\rangle .
\end{equation}%
The dispersion relation of the CRW-$c$ can be obtained from Eq.~(\ref{eq11}) by setting
the superscript $l=c$. Substituting the stationary eigenstate and the
Hamiltonian into the eigenequation $H_{\mathrm{cir,I}}\left\vert
E\right\rangle =E\left\vert E\right\rangle $, we can obtain the coupled
equations for the probability amplitudes as in Eqs.~(\ref{eq12})-(\ref{eq16}) but with Eq.~(\ref{eq15})
replaced by the following two equations
\begin{equation}
J_{a,2}u_{a}\left( 0\right) +J_{b,2}u_{b}\left( 0\right) +J_{c,2}u_{c}\left(
0\right) =\left( E-\Delta _{2}\right) u_{d2},
\end{equation}%
\begin{equation}
J_{c,2}u_{d2}-\xi _{c}u_{c}\left( 1\right) =Eu_{c}\left( 0\right) ,
\end{equation}%
and the subscript $l$ in Eq.~(\ref{eq16}) is replaced by $l=a,b,c$.

If a single photon with energy $E$ is incident from the infinity side of CRW-$l$%
, the interactions between cavity $l_{0}$ and $l_{0}^{\prime }$ ($%
l_{0},l_{0}^{\prime }=a_{0},b_{0},c_{0}$) mediated by two mechanical
modes will result in photon sacttering between different quantum channels.
The general expressions for the probability amplitudes in three quantum
channels ($l=a,b,c$) are given by ($j\geq 0$)
\begin{equation}\label{eq41}
u_{l}\left( j\right) =e^{-ik_{l}j}+s_{ll}e^{ik_{l}j},
\end{equation}%
\begin{equation}
u_{l^{\prime }}\left( j\right) =s_{l^{\prime }l}e^{ik_{l^{\prime }}j},
\end{equation}%
\begin{equation}\label{eq43}
u_{l^{\prime \prime }}\left( j\right) =s_{l^{\prime \prime
}l}e^{ik_{l^{\prime \prime }}j},
\end{equation}%
where $s_{l^{\prime }l}$ ($s_{l^{\prime \prime }l}$) denotes the scattering
amplitude from CRW-$l$ to CRW-$l^{\prime }$ (CRW-$l^{\prime \prime }$).
Substituting Eqs.~(\ref{eq41})-(\ref{eq43}) into the coupled equations for the
probability amplitudes, we can obtain the scattering matrix as%
\begin{equation}\label{eq44}
S=M^{-1}N,
\end{equation}%
with
\begin{equation}
S=\left(
\begin{array}{ccc}
s_{aa} & s_{ab} & s_{ac} \\
s_{ba} & s_{bb} & s_{bc} \\
s_{ca} & s_{cb} & s_{cc}%
\end{array}%
\right) ,
\end{equation}%
\begin{equation}
M=\left(
\begin{array}{ccc}
\xi _{a}^{\prime }e^{-ik_{a}^{\prime }} & J_{ab}e^{i\phi ^{\prime }} & J_{ca}
\\
J_{ab}e^{-i\phi ^{\prime }} & \xi _{b}^{\prime }e^{-ik_{b}^{\prime }} &
J_{bc} \\
J_{ca} & J_{bc} & \xi _{c}^{\prime }e^{-ik_{c}^{\prime }}%
\end{array}%
\right) ,
\end{equation}%
\begin{equation}
N=-\left(
\begin{array}{ccc}
\xi _{a}^{\prime }e^{ik_{a}^{\prime }} & J_{ab}e^{i\phi ^{\prime }} & J_{ca}
\\
J_{ab}e^{-i\phi ^{\prime }} & \xi _{b}^{\prime }e^{ik_{b}^{\prime }} & J_{bc}
\\
J_{ca} & J_{bc} & \xi _{c}^{\prime }e^{ik_{c}^{\prime }}%
\end{array}%
\right) ,
\end{equation}%
where the renormalized coupling strength $\xi _{l}^{\prime }$ and wave
number $k_{l}^{\prime }$ of the single photon in the CRW-$l$ are defined by
\begin{equation}\label{eq48}
\xi _{l}^{\prime }e^{ik_{l}^{\prime }}\equiv \xi _{l}e^{ik_{l}}+\Delta _{l}.
\end{equation}%
The effective coupling strengths $J_{ll^{\prime }}$, phase $\phi ^{\prime }$%
, and frequency shifts $\Delta _{l}$ induced by the two mechanical modes
are defined by%
\begin{equation}
J_{ab}e^{i\phi ^{\prime }}\equiv \frac{J_{a,1}J_{b,1}e^{i\phi }}{\left( E-\Delta
_{1}\right) }+\frac{J_{a,2}J_{b,2}}{\left( E-\Delta _{2}\right) },
\end{equation}%
\begin{equation}
J_{ca}\equiv \frac{J_{a,2}J_{c,2}}{\left( E-\Delta _{2}\right) },
\end{equation}%
\begin{equation}\label{eq51}
J_{bc}\equiv \frac{J_{b,2}J_{c,2}}{\left( E-\Delta _{2}\right) },
\end{equation}%
\begin{equation}
\Delta _{a} \equiv \frac{\left( J_{a,1}\right) ^{2}}{\left( E-\Delta _{1}\right) }+%
\frac{\left( J_{a,2}\right) ^{2}}{\left( E-\Delta _{2}\right) },
\end{equation}%
\begin{equation}
\Delta _{b} \equiv \frac{\left( J_{b,1}\right) ^{2}}{\left( E-\Delta _{1}\right) }+%
\frac{\left( J_{b,2}\right) ^{2}}{\left( E-\Delta _{2}\right) },
\end{equation}%
\begin{equation}
\Delta _{c} \equiv \frac{\left( J_{c,2}\right) ^{2}}{\left( E-\Delta _{2}\right) }.
\end{equation}

\subsection{Single-photon circulator}

\begin{figure}[tbp]
\includegraphics[bb=126 143 473 565, width=8.5 cm, clip]{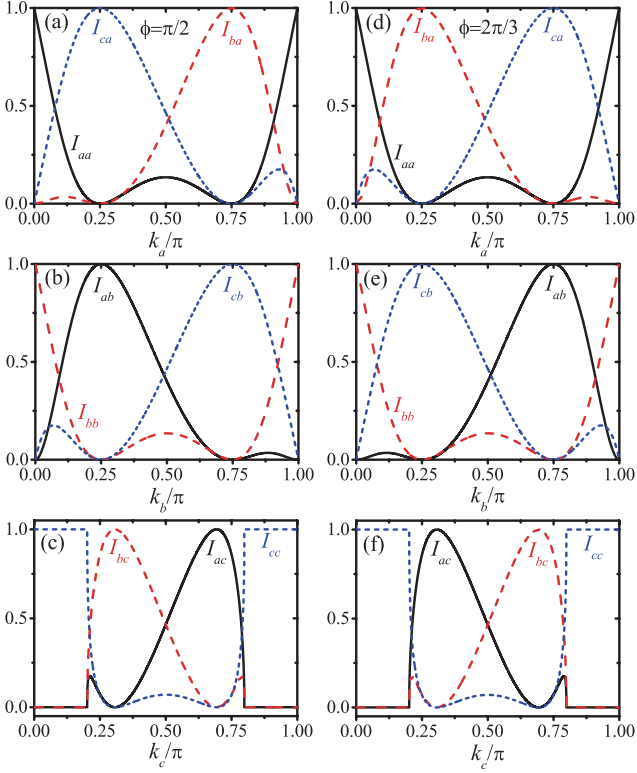}
\caption{(Color online) Scattering flows $I_{l^{\prime }l}$ ($l,l^{\prime
}=a,b,c$) as functions of the wave number $k_{l}/\protect\pi$ of a single
photon incident from CRW-$l$ for (a)-(c) $\protect\phi=\protect\pi/2$,
(d)-(f) $\protect\phi=3\protect\pi/2$. $J_{c,2}$ and $\protect\xi _{c}$ are obtained from Eqs.~(\ref{eq57}) and
(\ref{eq58}), and the other parameters are $\Delta
_{1}=\Delta _{2}=0$, $J_{a,1}=J_{b,1}=\protect\xi$, and $J_{a,2}=J_{b,2}=1.2%
\protect\xi$.}
\label{fig5}
\end{figure}

\begin{figure}[tbp]
\includegraphics[bb=122 175 465 600, width=8.5 cm, clip]{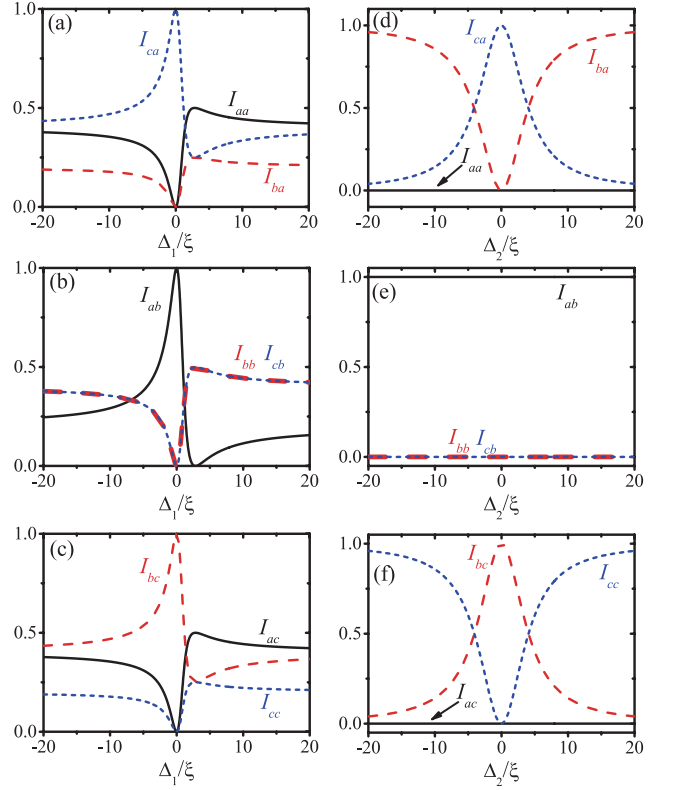}
\caption{(Color online) Scattering flows $I_{l^{\prime }l}$ ($l,l^{\prime
}=a,b,c$) are plotted as functions of (a)-(c) $\Delta _{1}/\protect\xi$ ($%
\Delta _{2}=0$) and (d)-(f) $\Delta _{2}/\protect\xi$ ($\Delta _{1}=0$) for $%
\protect\phi=\protect\pi/2$ and $k=\protect\pi/4$ ($k_{a}=k_{b}=k$, $k_{c}=\arccos[\xi\cos(k)/\xi_{c}]$). $J_{c,2} $ and $\protect\xi _{c}$ are obtained from Eqs.~(\ref{eq57}) and (\ref{eq58}), and the other parameters are $%
J_{a,1}=J_{b,1}=\protect\xi$ and $J_{a,2}=J_{b,2}=1.2\protect\xi$.}
\label{fig6}
\end{figure}

Let us give the optimal conditions for observing perfect circulators first.
A perfect circulator is obtained when we have $I_{ba}=I_{cb}=I_{ac}=1$ or $%
I_{ab}=I_{bc}=I_{ca}=1$ and the other scattering flows are equal to zero. For
the sake of simplicity, we assume that the detunings $\Delta _{1}=\Delta
_{2}=0$, the CRW-$a$ and CRW-$b$ have the same parameters (i.e. $\xi \equiv
\xi _{a}=\xi _{b}$, $k\equiv k_{a}=k_{b}$), and they are symmetrically
coupled to the two mechanical modes (i.e. $J_{1}\equiv J_{a,1}=J_{b,1}$, $%
J_{2}\equiv J_{a,2}=J_{b,2}$) with the coupling strength $J_{1}=\xi $. Based on
these assumptions, the perfect circulator appears with parameters satisfying the optimal conditions%
\begin{equation}\label{eq55}
\phi =\frac{\pi }{2}\quad \mathrm{or}\quad \frac{3\pi }{2},
\end{equation}%
\begin{equation}\label{eq56}
k=\frac{\pi }{4}\quad \mathrm{or}\quad \frac{3\pi }{4},
\end{equation}%
\begin{equation}\label{eq57}
\frac{J_{c,2}}{\xi }=\sqrt{\left( \frac{J_{2}}{\xi }\right) ^{4}+1},
\end{equation}%
\begin{equation}\label{eq58}
\xi _{c}=\left|\frac{\left(J_{bc}\right)^{2}}{\xi e^{-ik}+\Delta _{a}}-\Delta _{c}\right|,
\end{equation}%
where $J_{bc}$ has been given in Eq.~(\ref{eq51}).

In Fig.~\ref{fig5}, the scattering flows $I_{l^{\prime }l}$ ($l,l^{\prime
}=a,b,c$) are plotted as functions of the wave number $k_{l}/\protect\pi$ of a single
photon incident from CRW-$l$ for (a)-(c) $%
\phi =\pi /2$ and (d)-(f) $\phi =3\pi /2$. As shown in Figs.~\ref{fig5}%
(a)-(c), when $\phi =\pi /2$, we obtain that $I_{ba}=I_{cb}=I_{ac}=1$ and the
other scattering flows are equal to zero for the wave number $k=\pi /4$ ($k_{a}=k_{b}=k$, $k_{c}=\arccos[\xi\cos(k)/\xi_{c}]$), or obtain that
$I_{ab}=I_{bc}=I_{ca}=1$ and the other scattering flows are equal to zero for the wave number
$k=3\pi /4$ ($k_{a}=k_{b}=k$, $k_{c}=\arccos[\xi\cos(k)/\xi_{c}]$). As shown in Figs.~\ref{fig5}(d)-(f), when $%
\phi =3\pi /2$, we get $I_{ab}=I_{bc}=I_{ca}=1$ with the other zero scattering
flows for the wave number $k=\pi /4$ ($k_{a}=k_{b}=k$, $k_{c}=\arccos[\xi\cos(k)/\xi_{c}]$) or get $%
I_{ba}=I_{cb}=I_{ac}=1$ with the other zero scattering flows for the
wave number $k=3\pi /4$ ($k_{a}=k_{b}=k$, $k_{c}=\arccos[\xi\cos(k)/\xi_{c}]$). In other words, when $\phi =\pi /2$, the
signal is transferred from one CRW to another clockwise ($a\rightarrow
b\rightarrow c\rightarrow a$) for the wave number $k=3\pi /4$ ($k_{a}=k_{b}=k$, $k_{c}=\arccos[\xi\cos(k)/\xi_{c}]$) or
counterclockwise ($a\rightarrow c\rightarrow b\rightarrow a$)  for the wave number $%
k=\pi /4$ ($k_{a}=k_{b}=k$, $k_{c}=\arccos[\xi\cos(k)/\xi_{c}]$). In contrast, when $\phi =3\pi /2$, the signal is
transferred from one CRW to another counterclockwise ($a\rightarrow
c\rightarrow b\rightarrow a$) for the wave number $k=3\pi /4$ ($k_{a}=k_{b}=k$, $k_{c}=\arccos[\xi\cos(k)/\xi_{c}]$) or clockwise ($a\rightarrow b\rightarrow
c\rightarrow a$)  for the wave number $k=\pi /4$ ($k_{a}=k_{b}=k$, $k_{c}=\arccos[\xi\cos(k)/\xi_{c}]$).
Thus, we can reverse the direction of the circulator by tuning the phase from $\phi=\pi/2$ to $\phi=3\pi/2$.

Scattering flows $I_{l^{\prime }l}$ ($l,l^{\prime }=a,b,c$) as functions of
the detunings $\Delta _{1}/\xi $ and $\Delta _{2}/\xi $ are shown in Fig.~%
\ref{fig6}. Overall, the large detunings (both $\Delta _{1}$ and $\Delta _{2}
$) are destructive for the circulator. Similar to Fig.~\ref{fig3}(a), when
the detuning is tuned from $\Delta _{1}=0$ to $\Delta _{1}=2\sqrt{2}\xi $ as shown in Figs.~%
\ref{fig6}(a)-(c), the scattering flows $I_{ba}$ and $I_{ab}$ change from
($I_{ba}=0$, $I_{ab}=1$) to ($I_{ba}=0.25$, $I_{ab}=0$), i.e., the direction of the scattering flows
change from $b\rightarrow a$ to $a\rightarrow b$. What's more, when $%
I_{ab}=1$ and $I_{bb}=I_{cb}=0$ as shown in Fig.~\ref{fig3}(e), the
scattering flows $I_{ab}$, $I_{bb}$, and $I_{ab}$ remain constant for different
$\Delta _{2}$.

As shown in Eq.~(\ref{eq58}), i.e., $\xi _{c}\neq \xi $, the band widths of CRW-$a$ and CRW-$b$ are different
from the band width of CRW-$c$, and nonreciprocity ($I_{l^{\prime }l}\neq
I_{ll^{\prime }}$) can only be obtained in the overlap band regime between the
three CRWs. As shown in Fig.~\ref{fig5}(c) and \ref{fig5}(f), the single photon incident from the infinity side of CRW-$c$
will be reflected totally ($I_{cc}=1$) in the regimes of $0<k_{c}<\arccos(\xi/\xi_{c})$
and $\pi-\arccos(\xi/\xi_{c})<k_{c}<\pi $. Moreover, as shown in Eqs.~(\ref{eq55}) and (\ref{eq56}), we can have perfect
circulator only with wave number $k=\pi /4$ and $k=3\pi /4$ for $\phi =\pi /2$ or $3\pi
/2$. To improve tunability of the circulator, e.g., the perfect
circulator can be obtained with the wave number in a tunable regime, we
can use one more mechanical mode to connect the three CRWs and this is
the focus of the next section.

\section{Single-photon circulator with three mechanical modes}

\subsection{Theoretical model and scattering matrix}

\begin{figure}[tbp]
\includegraphics[bb=5 276 579 668, width=8.5 cm, clip]{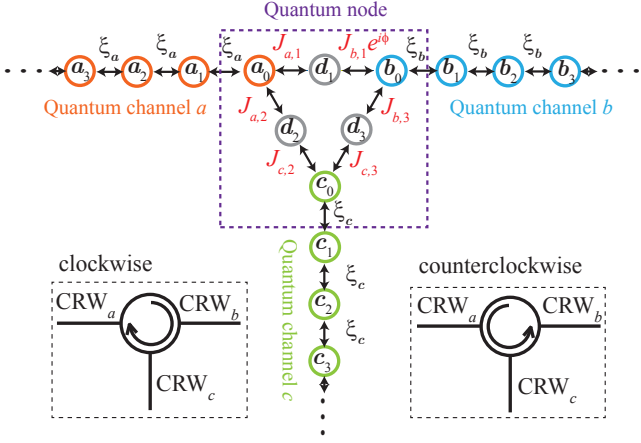}
\caption{(Color online) Schematic diagram of a T-shaped waveguide consisting
of three semi-infinite CRWs ($a_{j}$, $b_{j}$ and $c_{j}$ for $j\geq 0$)
coupled indirectly by three mechanical modes ($d_{1}$, $d_{2}$ and $d_{3}$).}
\label{fig7}
\end{figure}

In this section, we propose another single-photon circulator by a different
T-shaped waveguide, schematically shown in Fig.~\ref{fig4}, which is made up
of three semi-infinite CRWs coupled indirectedly by three mechanical modes
via optomechanical interactions. The T-shaped waveguide can be described by
the Hamiltonian%
\begin{equation}
H_{\mathrm{T,II}}=\sum_{l=a,b,c}H_{l}+H_{m}+H_{\mathrm{om,int}}+H_{\mathrm{dri}}
\end{equation}%
with the Hamiltonian $H_{l}$ for the CRWs given in Eq.~(\ref{eq2}), the Hamiltonian $%
H_{m}$ for three mechanical modes%
\begin{equation}
H_{m}=\omega _{1}d_{1}^{\dag }d_{1}+\omega _{2}d_{2}^{\dag }d_{2}+\omega
_{3}d_{3}^{\dag }d_{3},
\end{equation}%
the interactions $H_{\mathrm{om,int}}$ mediated by three mechanical modes
\begin{eqnarray}
H_{\mathrm{om,int}} &=&\left( g_{a,1}a_{0}^{\dag }a_{0}+g_{b,1}b_{0}^{\dag
}b_{0}\right) \left( d_{1}+d_{1}^{\dag }\right)   \notag \\
&&+\left( g_{a,2}a_{0}^{\dag }a_{0}+g_{c,2}c_{0}^{\dag }c_{0}\right) \left(
d_{2}+d_{2}^{\dag }\right)   \notag \\
&&+\left( g_{b,3}b_{0}^{\dag }b_{0}+g_{c,3}c_{0}^{\dag }c_{0}\right) \left(
d_{3}+d_{3}^{\dag }\right) ,
\end{eqnarray}%
and the externally driving terms%
\begin{eqnarray}
H_{\mathrm{dri}} &=&\sum_{i=1,2}a_{0}\Omega _{a,i}e^{i\omega
_{a,i}t}+\sum_{i=1,3}b_{0}\Omega _{b,i}e^{i\omega _{b,i}t}  \notag \\
&&+\sum_{i=2,3}c_{0}\Omega _{c,i}e^{i\omega _{c,i}t}+\mathrm{H.c.},
\end{eqnarray}%
where $\omega _{i}$ ($i=1,2,3$) is the resonant frequency of $i$th mechanical mode
with the bosonic annihilation (creation) operator $d_{i}$ ($d_{i}^{\dag }$).
$g_{l,i}$ with $l=a,b,c$ is the optomechanical coupling strength between cavity $l_{0}$ ($%
l_{0}=a_{0},b_{0},c_{0}$) and mechanical mode $d_{i}$ ($%
d_{i}=d_{1},d_{2},d_{3}$). Cavity $a_{0}$ ($b_{0}$, $c_{0}$) is driven by a
two-tone laser at frequencies $\omega _{a,1}=\omega _{a}-\omega _{1}+\Delta
_{a,1}$ and $\omega _{a,2}=\omega _{a}-\omega _{2}+\Delta _{a,2}$ ($\omega
_{b,1}=\omega _{b}-\omega _{1}+\Delta _{b,1}$ and $\omega _{b,3}=\omega
_{b}-\omega _{3}+\Delta _{b,3}$, $\omega _{c,2}=\omega _{c}-\omega _{2}+\Delta
_{c,2}$ and $\omega _{c,3}=\omega _{c}-\omega _{3}+\Delta _{c,3}$), and the amplitudes are $\Omega _{a,1}$ and $\Omega _{a,2}$ ($\Omega _{b,1}$ and $\Omega _{b,3}$, $\Omega _{c,2}$ and $\Omega _{c,3}$).
Here, we assume that the detunings satisfy the conditions $\Delta_{1}\equiv\Delta_{a,1}=\Delta_{b,1}$, $\Delta_{2}\equiv\Delta_{a,2}=\Delta_{c,2}$, and $\Delta_{3}\equiv\Delta_{b,3}=\Delta_{c,3}$.
Thus the operators for the
cavity modes can be rewritten as the sum of its quantum fluctuation operators
and classical mean values as $a_{j}\rightarrow a_{j}+\sum_{i=1,2}\alpha
_{j,i}^{a}e^{-i\omega _{a,i}t}$, $b_{j}\rightarrow b_{j}+\sum_{i=1,3}\alpha
_{j,i}^{b}e^{-i\omega _{b,i}t}$ and $c_{j}\rightarrow
c_{j}+\sum_{i=2,3}\alpha _{j,i}^{c}e^{-i\omega _{c,i}t}$, where the
classical amplitude $\alpha _{j,i}^{l}$ is determined by the amplitudes $%
\Omega _{l,i}$, the frequencies $\omega _{l,i}$, the damping
rates $\kappa _{l,j}$ of the cavities, and the mechanical damping rates $\gamma _{i}$.

To obtain a linearized Hamiltonian, similar to the previous assumptions that the external driving is
strong, i.e. $|\alpha _{0,i}^{l}|\gg 1$, the system works in the
resolved-sideband limit with respect to all three mechanical modes, i.e. $\min
\left\{ \omega _{i}\right\} \gg \max \left\{ \kappa _{l,j}\right\} $, and
the three mechanical modes are well separated in frequency, i.e. $\min
\left\{ \omega _{i},\left\vert \omega _{i}-\omega _{i^{\prime }\neq
i}\right\vert \right\} \gg \max \left\{ |g_{l,i}\alpha _{0,i}^{l}|,\gamma
_{i}\right\} $. After doing the standard linearization under the rotating-wave
approximation, in the rotating reference frame with respect to $H_{\mathrm{rot}%
}=\sum_{l=a,b,c}\sum_{j=0}^{+\infty }\omega _{l}l_{j}^{\dag
}l_{j}+\sum_{i=1,2,3}\left( \omega _{i}-\Delta _{i}\right) d_{i}^{\dag }d_{i}
$, the linearized Hamiltonian is obtained
\begin{equation}\label{eq63}
H_{\mathrm{cir,II}}=\sum_{l=a,b,c}H_{l}+H_{m}+H_{\mathrm{int,II}}
\end{equation}%
with the Hamiltonian $H_{l}$ of the CRWs given in Eq.~(\ref{eq6}), the Hamiltonian of
the mechanical modes
\begin{equation}
H_{m}=\Delta _{1}d_{1}^{\dag }d_{1}+\Delta _{2}d_{2}^{\dag }d_{2}+\Delta
_{3}d_{3}^{\dag }d_{3},
\end{equation}%
and the interaction terms
\begin{eqnarray}\label{eq65}
H_{\mathrm{int,II}} &=&J_{a,1}\left( a_{0}^{\dag }d_{1}+a_{0}d_{1}^{\dag
}\right)   \notag \\
&&+J_{b,1}\left( e^{-i\phi }b_{0}^{\dag }d_{1}+e^{i\phi }b_{0}d_{1}^{\dag
}\right)   \notag \\
&&+J_{a,2}\left( a_{0}^{\dag }d_{2}+a_{0}d_{2}^{\dag }\right)   \notag \\
&&+J_{c,2}\left( c_{0}^{\dag }d_{2}+c_{0}d_{2}^{\dag }\right)   \notag \\
&&+J_{b,3}\left( b_{0}^{\dag }d_{3}+b_{0}d_{3}^{\dag }\right)   \notag \\
&&+J_{c,3}\left( c_{0}^{\dag }d_{3}+c_{0}d_{3}^{\dag }\right) .
\end{eqnarray}%
Here $J_{l,i}e^{i\phi _{l,i}}=g_{l,i}\alpha _{0,i}^{l}$ with $l=a,b,c$ is the effective
optomechanical coupling strength between the cavity $l_{0}$ ($%
l_{0}=a_{0},b_{0},c_{0}$) and mechanical mode $d_{i}$ ($%
d_{i}=d_{1},d_{2},d_{3}$) with real strength $J_{l,i}=\left\vert
g_{l,i}\alpha _{0,i}^{l}\right\vert $ and phase $\phi _{l,i}$. As only the total phase $\phi =\phi _{a,1}+\phi _{a,2}+\phi
_{b,1}+\phi _{b,3}+\phi _{c,2}+\phi _{c,3}$ among them has physical
effects, without loss of generality, $\phi $ is only kept in the terms of $%
b_{0}d_{1}^{\dag }$ and $b_{0}^{\dag }d_{1}$ in Eq.~(\ref{eq65}) and the following
derivation. Similarly, $J_{l,i}$ and $\phi $ can be
controlled dynamically by tuning the strengths and phases of the external driving
fields.

The stationary eigenstate of a single photon scattering in the T-shaped
waveguide with three mechanical modes is given by
\begin{equation}\label{eq66}
\left\vert E\right\rangle =\sum_{l=a,b,c}\sum_{j=0}^{+\infty }u_{l}\left(
j\right) l_{j}^{\dag }\left\vert 0\right\rangle
+\sum_{i=1}^{3}u_{di}d_{i}^{\dag }\left\vert 0\right\rangle.
\end{equation}%
Substituting the stationary eigenstate in Eq.~(\ref{eq66}) and the Hamiltonian in Eq.~(\ref{eq63}) into the
eigenequation $H_{\mathrm{cir,II}}\left\vert E\right\rangle =E\left\vert
E\right\rangle $, we can obtain the coupled equations for the probability
amplitudes as
\begin{equation}
J_{a,1}u_{d1}+J_{a,2}u_{d2}-\xi _{a}u_{a}\left( 1\right) =Eu_{a}\left(
0\right) ,
\end{equation}%
\begin{equation}
J_{b,1}e^{-i\phi }u_{d1}+J_{b,3}u_{d3}-\xi _{b}u_{b}\left( 1\right)
=Eu_{b}\left( 0\right) ,
\end{equation}%
\begin{equation}
J_{c,2}u_{d2}+J_{c,3}u_{d3}-\xi _{c}u_{c}\left( 1\right) =Eu_{c}\left(
0\right) ,
\end{equation}%
\begin{equation}
J_{a,1}u_{a}\left( 0\right) +J_{b,1}e^{i\phi }u_{b}\left( 0\right) =\left(
E-\Delta _{1}\right) u_{d1},
\end{equation}%
\begin{equation}
J_{a,2}u_{a}\left( 0\right) +J_{c,2}u_{c}\left( 0\right) =\left( E-\Delta
_{2}\right) u_{d2},
\end{equation}%
\begin{equation}
J_{b,3}u_{b}\left( 0\right) +J_{c,3}u_{c}\left( 0\right) =\left( E-\Delta
_{3}\right) u_{d3},
\end{equation}%
and the same equation as Eq.~(\ref{eq16}) with the subscript $l=a,b,c$.

If a single photon with energy $E$ is incident from the infinity side of CRW-$l$%
, following similar steps given in Section II, we can obtain the scattering
matrix given in Eqs.~(\ref{eq44})-(\ref{eq48}) with the effective coupling strengths $%
J_{ll^{\prime }}$ , phase $\phi ^{\prime }$, and frequency shifts $\Delta
_{l}$ induced by the three mechanical modes redefined by
\begin{equation}\label{eq73}
J_{ab}e^{i\phi ^{\prime }}\equiv\frac{J_{a,1}J_{b,1}}{\left( E-\Delta _{1}\right)
}e^{i\phi },
\end{equation}%
\begin{equation}
J_{ca}\equiv\frac{J_{a,2}J_{c,2}}{\left( E-\Delta_{2}\right) },
\end{equation}%
\begin{equation}
J_{bc}\equiv\frac{J_{b,3}J_{c,3}}{\left( E-\Delta_{3}\right) },
\end{equation}%
\begin{equation}
\Delta _{a}\equiv\frac{\left( J_{a,1}\right) ^{2}}{\left( E-\Delta_{1}\right) }+%
\frac{\left( J_{a,2}\right) ^{2}}{\left( E-\Delta_{2}\right) },
\end{equation}%
\begin{equation}
\Delta _{b}\equiv\frac{\left( J_{b,1}\right) ^{2}}{\left( E-\Delta_{1}\right) }+%
\frac{\left( J_{b,3}\right) ^{2}}{\left( E-\Delta_{3}\right) },
\end{equation}%
\begin{equation}\label{eq78}
\Delta _{c}\equiv\frac{\left( J_{c,2}\right) ^{2}}{\left( E-\Delta_{2}\right) }+%
\frac{\left( J_{c,3}\right) ^{2}}{\left( E-\Delta_{3}\right) }.
\end{equation}

\subsection{Single-photon circulator}

\begin{figure}[tbp]
\includegraphics[bb=114 220 458 666, width=8.5 cm, clip]{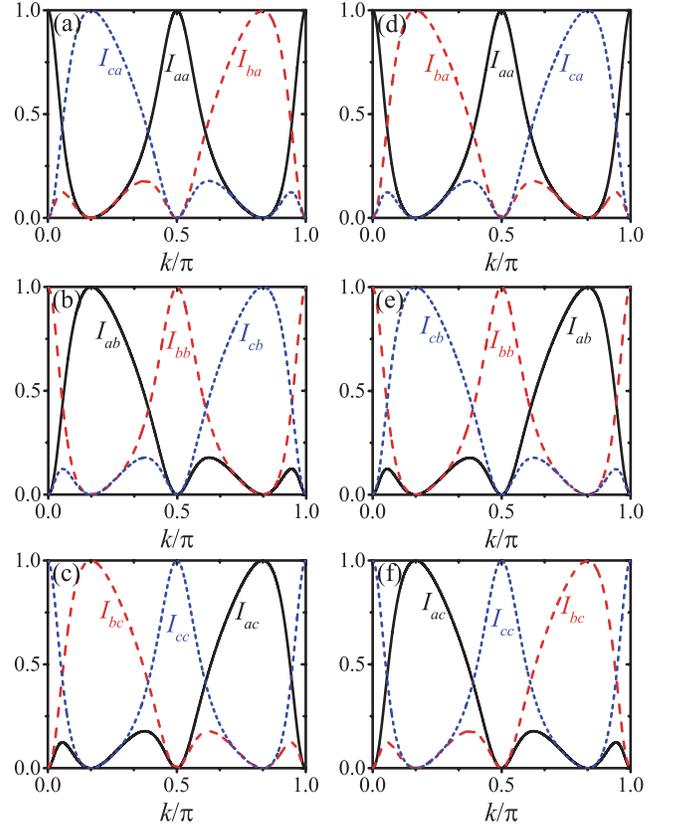}
\caption{(Color online) Scattering flows $I_{l^{\prime }l}$ ($l,l^{\prime
}=a,b,c$) as functions of the wave number $k/\protect\pi$ for $\protect\phi=\protect\pi/3$ in (a)-(c) and $\protect\phi=5\protect\pi/3$ in (d)-(f). The
other parameters are $\protect\xi _{c}=\protect\xi$, $k _{c}=k$, $\Delta
_{1}=\Delta _{2}=\Delta _{3}=0$, $J_{1}=J_{2}=J_{3}=J$, and $J$ is obtained from Eq.~(\ref{eq79}).}
\label{fig8}
\end{figure}

\begin{figure}[tbp]
\includegraphics[bb=113 249 455 694, width=8.5 cm, clip]{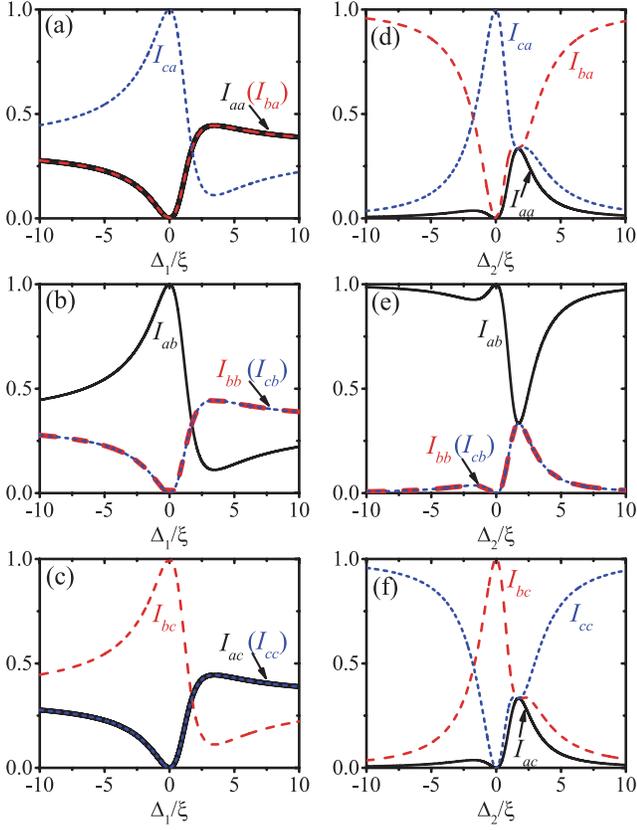}
\caption{(Color online) Scattering flows $I_{l^{\prime }l}$ ($l,l^{\prime
}=a,b,c$) are plotted as functions of (a)-(c) $\Delta _{1}/\protect\xi$ ($%
\Delta _{2}=\Delta _{3}=0$) and (d)-(f) $\Delta _{2}/\protect\xi$ ($\Delta
_{1}=0,\Delta _{3}=\Delta _{2}$) for $\protect\phi=\protect\pi/3$ and $k=0.5236$
[obtained from Eq.~(\ref{eq80})]. The other parameters are $\protect\xi _{c}=\protect%
\xi $, $k _{c}=k$, $J_{1}=J_{2}=J_{3}=J$, and $J$ is obtained
from Eq.~(\ref{eq79}).}
\label{fig9}
\end{figure}

\begin{figure}[tbp]
\includegraphics[bb=120 163 467 624, width=8.5 cm, clip]{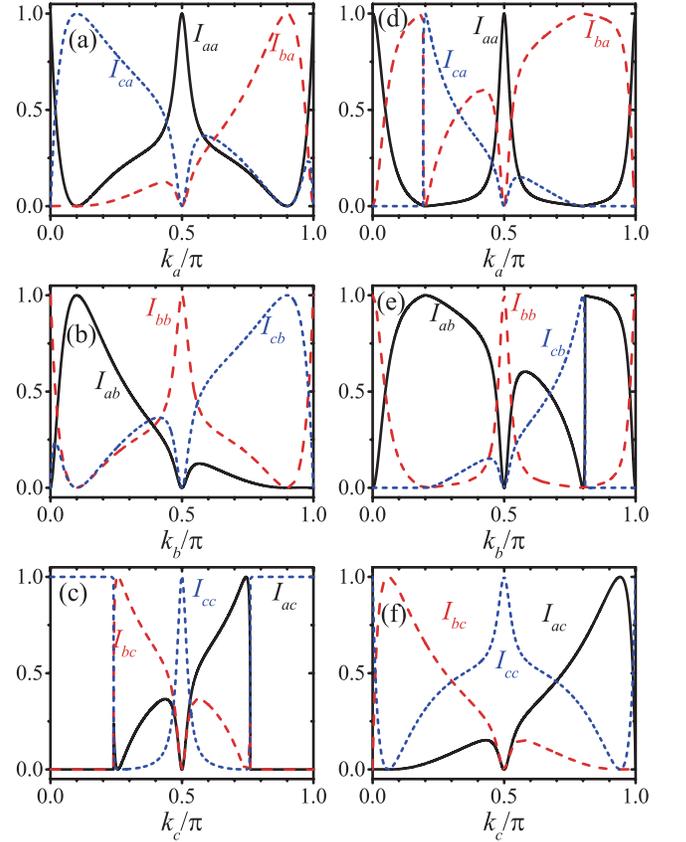}
\caption{(Color online) Scattering flows $I_{l^{\prime }l}$ ($l,l^{\prime
}=a,b,c$) are plotted as functions of the wave number $k_{l}/\protect\pi$ of a single
photon incident from CRW-$l$ for the perfect circulator appearing
at (a)-(c) $k=0.1\protect\pi$ and $0.9\protect\pi$, (d)-(f) $k=0.2\protect%
\pi $ and $0.8\protect\pi$. $J_{i}$ ($i=1,2,3$) and $\protect\xi_{c}$ are obtained from Eqs.~(\ref{eq83})-(\ref{eq86}), and the other parameters are $\protect\phi=\pi/2$ and $\Delta _{1}=\Delta _{2}=\Delta _{3}=0$.}
\label{fig10}
\end{figure}

The optimal conditions to obtain a perfect circulator can be derived
analytically from Eqs.~(\ref{eq44})-(\ref{eq48}) [with $J_{ll^{\prime }}$ , $\phi ^{\prime }$, and $\Delta
_{l}$ defined in Eqs.~(\ref{eq73})-(\ref{eq78})] by setting $I_{ba}=I_{cb}=I_{ac}=1$ or $%
I_{ab}=I_{bc}=I_{ca}=1$ and the other zero scattering flows. For
the sake of simplicity, we assume that the detunings $\Delta _{1}=\Delta
_{2}=\Delta _{3}=0$, CRW-$a$ and CRW-$b$ have the same parameters (i.e. $%
\xi \equiv \xi _{a}=\xi _{b}$, $k\equiv k_{a}=k_{b}$) with the coupling
strength $J_{1}\equiv J_{a,1}=J_{b,1}$, $J_{2}\equiv J_{a,2}=J_{b,3}$, $%
J_{3}\equiv J_{c,2}=J_{c,3}$. If $\xi _{c}=\xi $ and $J\equiv J_{1}=J_{2}=J_{3}$, the perfect circulator can be obtained
when the parameters satisfy the conditions
\begin{equation}\label{eq79}
\frac{J^{2}}{\xi ^{2}}=\frac{2\left( 2-\cos \phi \right) }{\left( 5-4\cos
\phi \right) },
\end{equation}%
\begin{eqnarray}\label{eq80}
k &=&\frac{1}{2}\arcsin \left\vert \frac{4\sin \phi -\sin 2\phi }{5-4\cos
\phi }\right\vert \text{ },  \notag \\
&&\mathrm{or} \quad \text{ }\pi -\frac{1}{2}\arcsin \left\vert \frac{4\sin \phi
-\sin 2\phi }{5-4\cos \phi }\right\vert .
\end{eqnarray}%
If $\xi _{c}\neq \xi $ and $J_{1}\neq J_{2}\neq J_{3}$, the perfect circulator can also be obtained but the conditions are changed to
\begin{equation}
\phi =\frac{\pi }{2}\quad \mathrm{{or}\quad }\frac{3\pi }{2},
\end{equation}%
\begin{equation}
k\in \left( 0,\frac{\pi }{4}\right) \cup \left( \frac{3\pi }{4},\pi \right) ,
\end{equation}%
and the optimal coupling strengths are given by
\begin{equation}\label{eq83}
\frac{J_{1}}{\xi }=\sqrt{\left\vert \sin 2k\right\vert },
\end{equation}%
\begin{equation}
\frac{J_{2}}{\xi }=\sqrt{2\cos ^{2}k-\left\vert \sin 2k\right\vert },
\end{equation}%
\begin{equation}
\frac{J_{3}}{\xi }=\left\vert \cos k\right\vert ,
\end{equation}%
\begin{equation}\label{eq86}
\frac{\xi _{c}}{\xi }=\left\vert \frac{\cos k}{\cos \left[ \arctan \left(
\frac{2\cos ^{2}k-\left\vert \sin 2k\right\vert }{4\left\vert \cos
k\right\vert \sin k}\right) \right] }\right\vert .
\end{equation}%
From Eqs.~(\ref{eq83}) and (\ref{eq86}), if the wave number $k$ to obtain optimal circulator
satisfies the equation
\begin{equation}\label{eq87}
2\cos ^{2}k-\left\vert \sin 2k\right\vert =4\sin ^{2}k,
\end{equation}%
we obtain $\xi =\xi _{c}$ and $J_{1}=J_{2}=J_{3}$, and this is consistent
with the condition given in Eq.~(\ref{eq80}) for $\phi =\pi/2$.

In Fig.~\ref{fig8}, the scattering flows $I_{l^{\prime }l}$ ($l,l^{\prime
}=a,b,c$) are plotted as functions of the wave number $k/\pi $ for (a)-(c) $%
\phi =\pi /3$, (d)-(f) $\phi =5\pi /3$ with $\xi
_{c}=\xi $ and $J\equiv J_{1}=J_{2}=J_{3}$. As shown in Figs.~\ref{fig8}%
(a)-(c), when $\phi =\pi /3$, we obtain that $I_{ab}=I_{bc}=I_{ca}=1$ and the
other scattering flows are equal to zero for wave number $k=\pi /6$, or obtain $%
I_{ba}=I_{cb}=I_{ac}=1$ and the other scattering flows are equal to zero
for wave number $k=5\pi /6$. As shown in Figs.~\ref{fig8}(d)-(f), when $\phi =5\pi /3$,
we get $I_{ba}=I_{cb}=I_{ac}=1$ with the other zero scattering flows
for wave number $k=\pi /6$, or get $I_{ab}=I_{bc}=I_{ca}=1$ with the
other zero scattering flows for wave number $k=5\pi /6$. In other
words, when $\phi =\pi /3$, the signal is transferred from one CRW to
another counterclockwise ($a\rightarrow c\rightarrow b\rightarrow a$) for wave number $%
k=\pi /6$ or clockwise ($a\rightarrow b\rightarrow c\rightarrow a$)
for wave number $k=5\pi /6$. In contrast, when $\phi =5\pi /3$, the signal
is transferred from one CRW to another clockwise ($%
a\rightarrow b\rightarrow c\rightarrow a$) for wave number $k=\pi /6$ or counterclockwise ($a\rightarrow
c\rightarrow b\rightarrow a$) for wave number $k=5\pi /6$.
In other words, we can tune the phase from $\phi$ to ($\pi-\phi$) to reverse the direction of the circulator.

The effects of the detunings on the single-photon transmission can be seen from Fig.~\ref{fig9},
where the scattering flows $I_{l^{\prime }l}$ ($l,l^{\prime }=a,b,c$) are plotted as functions of
the detunings (a)-(c) $\Delta _{1}/\xi $ ($\Delta _{2}=\Delta _{3}=0$) and
(d)-(f) $\Delta _{2}/\xi $ ($\Delta _{1}=0$ and $\Delta _{3}=\Delta _{2}$).
It is interesting that, we obtain $I_{l^{\prime
}l}=1/3$ ($l,l^{\prime }=a,b,c$) with the detunings $\Delta _{1}=\sqrt{3}\xi
$ ($\Delta _{2}=\Delta _{3}=0$) or $\Delta _{3}=\Delta _{2}=\sqrt{3}\xi $ ($%
\Delta _{1}=0$). This interesting phenomenon can be used to design three-way
single-photon beam splitter.

If we have different coupling strengths $\xi_{c}\neq \xi $, the band widths of CRW-$a$ and CRW-$b$ are different
from the band width of CRW-$c$, and nonreciprocity ($I_{ll^{\prime }}\neq
I_{l^{\prime }l}$) can only be obtained in the overlap band regime among the three CRWs.
In Fig.~\ref{fig10}, the scattering flows $I_{l^{\prime }l}$ ($l,l^{\prime
}=a,b,c$) are plotted as functions of the wave number $k_{l}/\pi $ of a
single photon incident from CRW-$l$ with different coupling strengths $\xi
_{c}\neq \xi $. Different from the case of T-shaped waveguide with two
mechanical modes as discussed in Sec.~III, we can have perfect circulator
with wave number $k\in \left( 0,\pi/4\right) \cup \left( 3\pi/4,\pi \right) $.
From Eq.~(\ref{eq87}) [or Eq.~(\ref{eq80}) for $\phi =\pi/2$], to make the perfect
circulator behavior appearing at $k=0.1476\pi$ and $0.8524\pi$, we should choose $\xi
_{c}=\xi $ and $J_{1}=J_{2}=J_{3}$. As shown in Figs.~\ref{fig10}(a)-(c),
when the perfect circulator appears at $k=0.1\pi$ ($<0.1476\pi$) and $0.9\pi$ ($>0.8524\pi$),
we have $\xi _{c}>\xi $. In Figs.~\ref{fig10}(d)-(f), when the perfect
circulator behavior appears at $k=0.2\pi$ ($>0.1476\pi$) and $0.8\pi$ ($<0.8524\pi$), we have $%
\xi _{c}<\xi $.

\section{Conclusions}

In summary, we have studied the nonreciprocal single-photon frequency
conversion in multiple CRWs, which are coupled indirectly by optomechanical interactions with two
nondegenerate mechanical modes. We have demonstrated that the frequency of a
single photon can be converted nonreciprocally in two CRWs. Moreover, two
different single-photon circulators are proposed in the T-shaped waveguides with two or three mechanical modes.
The optomechanical systems (or mechanical modes) offers the possibility to enable nonreciprocal frequency transduction between two CRWs with distinctively different frequencies, and they allow for dynamic control of the direction of frequency conversion by tuning the phases of external driving lasers.
All the proposals can be applied to integrate devices with different frequencies and simplify the construction of hybrid quantum
networks.

For simplicity, in this work we do not consider the dissipative effects of
the cavities. However, in reality, all the optical or microwave cavities in
the CRWs interact with the environment, resulting in the reduction of the
nonreciprocal single-photon frequency conversing efficiency. To lower this
effect, we should enhance both the coupling strength between two
nearest neighboring cavities in the CRWs and the effective optomechanical
coupling strengths. The photon hopping can be enhanced by decreasing the
distance between neighboring cavities, and one of the most common ways to
enhance effective optomechanical coupling strengths is to increase the
powers of the external driving fields. In this case, the frequencies of the
mechanical modes we choose must be high enough to ensure that the
rotating-wave approximation is valid in the derivations. Thus, the
microwave-frequency mechanical modes coupled to superconducting quantum
circuit~\cite{OConnellNat10} and optomechanical crystal~\cite%
{Safavi-NaeiniPRL14} are suitable to realize our nonreciprocal single-photon
frequency converters.

\vskip 2pc \leftline{\bf Acknowledgement}

X.-W.X. is supported by the National Natural Science Foundation of China
(NSFC) under Grant No.11604096 and the Startup Foundation for Doctors of
East China Jiaotong University under Grant No. 26541059. A.-X.C. is
supported by NSFC under Grant No. 11365009. Y.L. is supported by NSFC under
Grant No. 11421063. Y.L. and Y.-X.L. are supported by the National Basic
Research Program of China (973 Program) under Grant No. 2014CB921400.
Y.-X.L. is also supported by NSFC under Grants No. 61328502 and 61025022,
the Tsinghua University Initiative Scientific Research Program, and the
Tsinghua National Laboratory for Information Science and Technology (TNList)
Cross-Discipline Foundation.

\bibliographystyle{apsrev}
\bibliography{ref}

\end{document}